\documentclass[preprint,preprintnumbers,amsmath,amssymb,showkeys]{revtex4}

\usepackage{graphicx}
\usepackage{dcolumn}
\usepackage{bm}

\begin{document}

\title{Fluctuation-induced Non-equilibrium Transition in a Liquid-Crystal Metastable System}

\author{I. L. Ho}
\email{sunta.ho@msa.hinet.net}
\affiliation{Institute of Physics, Academia Sinica, Taipei, Taiwan, R.O.C\\}

\date{\today}

\begin{abstract}
The research herein studies the Langevin dynamics allowing for an
exchange of energy between liquid crystals and the thermal environment.
This dynamics leads to fluctuation and dissipation behaviors in the
motions of liquid crystals, and therefore drives the system toward
non-equilibrium evolutional processes. In particular, for the
operations of liquid-crystal metastable systems, the fluctuation could
allow an excitation (non-equilibrium) transition against energy
barriers to the globally-stable state. Implemented with an actual
case of liquid crystal $\pi$ configuration, this work statistically studies the
non-equilibrium metastable transitions
and shows the dependence of the transition-time on the
correlations (of fluctuations).

\end{abstract}

\maketitle

\section{INTRODUCTION}
In continuum models of liquid crystals \cite{lc1,lc2,lc3,lc4}, which
are based on phenomenological elastic constants and hydrodynamic
transport coefficients, the equations of motion for the liquid
crystal director are obtained by minimizing the free energy of the
system. However, in liquid-crystal metastable systems \cite{m1}, the
energy-minimization process probably causes the liquid crystal state
to be in a nearby local minimum of energy (weakly stable state) and
prohibits the barrier transition to the observable global minimum
state (strongly stable state). This process could result in a deviation
between theoretical predictions and observations in the metastable systems.
Accordingly, this
paper considers the non-equilibrium Langevin dynamics \cite{th1,th2,th3} which
allows an energy exchange between liquid
crystals and the thermal environment, and introduces fluctuation $\xi$
(absorbing energy) and dissipation $\propto\gamma$ (emitting energy)
behaviors associated with the motions of liquid crystals. In
principle, this is related to the fundamental fluctuation-dissipation relation
$\langle\xi(t_{1})\xi(t_{0})\rangle\propto \gamma(t_{1}-t_{0})$, and exhibits noticeable influences
in some liquid crystal systems. For instances, Barbero indicated that thermal fluctuations
can lead to a re-normalization of the anchoring energy of nematic liquid crystals \cite{fl1}.
Galatola studied the Freedericksz transitions in nematic liquid crystals
by thermal fluctuation approach \cite{fl2}. Kelly showed the director fluctuations in nematic droplet
with spherical harmonics method \cite{fl3}. Ruhwandl presented the topological defects
around a spherical colloid particle by Monte Carlo simulation \cite{fl4}.
Rehberg observed the spatial and temporal correlations of thermal fluctuations in liquid crystal
systems \cite{fl5}. Maclennan indicated a direct observation on the fluctuation of the liquid
crystal director in his work \cite{fl6}.
In this work, we mainly introduce the influences of thermal fluctuations by Langevin dynamics to study the
complete metastable transitions $V\rightarrow (T,H)\rightarrow H$ in liquid-crystal $\pi$
configurations \cite{OCB1}, which still remains indefinite in real three-dimensional systems as we know.

Based on Frank-Oseen elastic theory\cite{Q1}, we consider the spatiotemporal-correlated
fluctuations $\xi$ as well as the
corresponding dissipations $\propto\gamma$ for the liquid crystal directors in Langevin dynamics,
since (I) the common non-correlated fluctuations $\xi_{n}$ \cite{lag1} conditioned
by $\langle\xi_{n}(t_{1})\xi_{n}(t_{0})\rangle\propto \gamma k_{B}T\delta(t_{1}-t_{0})$ will cause
a non-physical/divergent quantity in ultra-small time scale $t_{1}-t_{0}\approx 0$, while (II)
the spatiotemporal-correlated fluctuations $\xi$ offers a more real algorithm to describe the experimental
observations \cite{fl5}.
Definitely, we consider the fluctuations
 $\xi$ related to an exponential correlation behavior
$\langle\xi(r_{1},t_{1})\xi(r_{0},t_{0})\rangle\propto \gamma k_{B}T e^{-\frac{|t_{1}-t_{0}|}{\tau}}e^{-\frac{|r_{1}-r_{0}|}{\lambda_{z}}}$
in the 1+1 (spatial+temporal) dimensions via the Ornstein-Uhlenbeck
process\cite{OU1}, and thereby characterize the fluctuation behavior by
correlation time $\tau$ and length $\lambda$ as experimental observations \cite{fl5}.
Implemented with an actual case
of the metastable systems in the liquid-crystal $\pi$ configuration
\cite{OCB1}, this paper fulfills a statistical numerical analysis
over an ensemble of systems with 1000 identical iterations, and
elucidates the non-equilibrium metastable transitions.

This article is organized as follows. Section 2 reviews Frank-Ossen
elastic free energy using Q tensor representation\cite{Q1}, and
structures the fundamental equation of motion for liquid crystal
directors. Section 3 further introduces the Ornstein-Uhlenbeck
process \cite{OU1} for the Langevin dynamics to consider an actual
spatiotemporal-correlated fluctuations $\xi(r,t)$.
An update formula of directors for the numerical calculation is then
identified. Section 4 numerically analyzes the influences of fluctuations and dissipations
 in the liquid crystal $\pi$ configuration, and depicts
the associated non-equilibrium metastable transitions, i.e., about how a initial
$V$ state transit through the locally-stable $T$ and finally into the globally-stable $H$ state
in the real three-dimensional systems. The main conclusions are
then discussed in section 5. Two more experiments as well as the corresponding theoretical works
are appendixed and show the influences of fluctuations and dissipations even in usual
cells of Liquid-Crystal Displays (LCDs).

\section{Modeling Liquid Crystal Director}

In elastic continuum theory, the local aligned sample may be regarded
as a single liquid crystal, in which the molecules are on the
average aligned along the direction defined by a director
$\vec{n}$. The Frank-Oseen strain free energy density using tensor
representation for liquid crystal directors can then be described
by\cite{Q1}:

\begin{eqnarray}
f_{s} &=&\frac{1}{12}(K_{33}-K_{11}+3K_{22})\frac{G_{1}^{(2)}}{S^{2}}+\frac{1%
}{2}(K_{11}-K_{22})\frac{G_{2}^{(2)}}{S^{2}}  \nonumber \\
&&+\frac{1}{4}(K_{33}-K_{11})\frac{G_{6}^{(3)}}{S^{3}}+q_{0}K_{22}\frac{%
G_{4}^{(2)}}{S^{2}},
\end{eqnarray}
with
\begin{eqnarray}
G_{1}^{(2)} &=&Q_{jk,l}Q_{jk,l}\text{ \ \ \ }G_{4}^{(2)}=\varepsilon
_{jkl}Q_{jm}Q_{km,l} \\
G_{2}^{(2)} &=&Q_{jk,k}Q_{jl,l}\text{ \ \ \ }%
G_{6}^{(3)}=Q_{jk}Q_{lm,j}Q_{lm,k}
\end{eqnarray}
in which
\begin{eqnarray}
Q_{jk} &=&S\left( n_{j}n_{k}-\frac{1}{3}\delta _{jk}\right) , \\
Q_{jk,l} &=&\frac{\partial Q_{jk}}{\partial l},...\text{ \ \
}j,k,l\in \{x,y,z\}.
\end{eqnarray}
Here, $K_{11}$, $K_{22} $, and $K_{33}$ are the splay, twist, and
bend elastic constants, respectively. $q_{0}=2\pi/p_{0}$ associates
the pitch $p_{0}$ of chiral liquid crystals, e.g. Cholesterics. Term
Q is the tensor order parameter, and $\vec{n}$ is the liquid crystal
director with unit length. $S$ is the scalar order parameter, and
indicates the ideal isotropic and nematic phases by the values $S=0$ and
$S=1$, respectively. $\varepsilon _{jkl}$ is the Levi-Civita symbol.
$\delta _{jk}$ represent the Kronecker delta. Moreover, the
convention of summing over repeated indices is used herein. The
electric free energy density can be expressed in terms of the Q
tensor as
\begin{equation}
f_{e}=\frac{1}{2}\epsilon _{0}\overline{\epsilon }V_{,j}V_{,j}+\frac{1}{2}%
\epsilon _{0}\bigtriangleup \epsilon V_{,j}V_{,k}\frac{Q_{jk}}{S},
\end{equation}
with
\begin{eqnarray}
\overline{\epsilon } &=&\frac{2\epsilon _{\perp }+\epsilon _{\parallel }}{3},%
\text{ \ \ }\bigtriangleup \epsilon =\epsilon _{\parallel }-\epsilon
_{\perp
}, \\
V_{,j} &=&\frac{\partial V}{\partial j},...\text{ \ \ }j\in
\{x,y,z\}.
\end{eqnarray}
Here, $\mathbf{\epsilon}$ is the LC dielectric tensor with
elements $\epsilon _{\perp }$ and $\epsilon _{\parallel }$ for dielectric
coefficients perpendicular and parallel to the LC director.

To solve equilibrium states of liquid crystal directors with applied
potentials, the system free energy, which is obtained by integrating
the Gibbs free energy density $f_{g}=f_{s}-f_{e}$ over the volume,
is minimized. Accordingly, the Euler-Lagrange equations
for electric potentials and director components under condition
$\left\vert \vec{n}\right\vert =1$ can thus be obtained

\begin{eqnarray}
0 &=&-\left[ f_{g}\right] _{V}=\partial _{i}%
\left[ (\varepsilon _{\bot }\delta _{ij}+(\varepsilon _{\Vert }-\varepsilon
_{\bot })n_{i}n_{j})\partial _{j}V\right] , \label{solvV} \\
0 &=&-\left[ f_{g}\right] _{n_{i}}+\lambda n_{i}\text{ \ \ }i\in
\{x,y,z\}, \label{solvn}
\end{eqnarray}
Here, $\lambda$ is the Lagrange multiplier for constraint
$\left\vert \vec{n}\right\vert =1$. By applying the chain rule
$\left[ f_{g}\right] _{n_{i}}=\left[ f_{g}\right]
_{Q_{jk}}\frac{\partial Q_{jk}}{\partial n_{i}}=\left[ f_{g}\right]
_{Q_{jk}}\cdot S\left( n_{j}\delta _{ki}+n_{k}\delta _{ji}\right) $,
we obtain the alternative definitions in tensor representation for Equation (\ref{solvn})
\begin{eqnarray}
\left[ f_{g}\right] _{n_{x}} &=&2n_{x}S\left[ f_{g}\right]
_{Q_{xx}}+n_{y}S\left( \left[ f_{g}\right] _{Q_{xy}}+\left[
f_{g}\right]
_{Q_{yx}}\right)   \nonumber \\
&&+n_{z}S\left( \left[ f_{g}\right] _{Q_{xz}}+\left[ f_{g}\right]
_{Q_{zx}}\right) , \\
\left[ f_{g}\right] _{n_{y}} &=&2n_{y}S\left[ f_{g}\right]
_{Q_{yy}}+n_{x}S\left( \left[ f_{g}\right] _{Q_{xy}}+\left[
f_{g}\right]
_{Q_{yx}}\right)   \nonumber \\
&&+n_{z}S\left( \left[ f_{g}\right] _{Q_{yz}}+\left[ f_{g}\right]
_{Q_{zy}}\right) , \\
\left[ f_{g}\right] _{n_{z}} &=&2n_{z}S\left[ f_{g}\right]
_{Q_{zz}}+n_{x}S\left( \left[ f_{g}\right] _{Q_{xz}}+\left[
f_{g}\right]
_{Q_{zx}}\right)   \nonumber \\
&&+n_{y}S\left( \left[ f_{g}\right] _{Q_{yz}}+\left[ f_{g}\right]
_{Q_{zy}}\right)
\end{eqnarray}

Solving Equations (\ref{solvV}) and (\ref{solvn}) simultaneously, the potential
distribution and the director configuration for equilibrium status
can now be evaluated. It is noted that, in general, $\left[
f_{g}\right] _{Q_{jk}}$ is not equal to $\left[ f_{g}\right]
_{Q_{kj}}$ for cases of $j\neq k$.
For the analyses of dynamic time evolutions on the liquid-crystal directors, however,
we consider the time-dependent/non-equilibrium Langevin equation.
In the Langevin dynamics which includes the stochastic torque (fluctuation) $\sigma\zeta$
as well as the dissipation torque $\kappa$ for directors,
the general rotational Langevin equation under a external torque $F$ can be read as \cite{lag1}:
\begin{equation}
\kappa\frac{\partial }{\partial t}n_{i}=-F_{i}+\sigma\zeta ,
\label{normallageq}
\end{equation}
Here the constant $\sigma$ is related to $\kappa$ by Einstein's relation
$\sigma=\sqrt{2\kappa k_{B}T}$, and $\zeta$ is the normalized white noise with dimension
$s^{-1/2}$. For the liquid-crystal system, to simply evaluate the dissipative torque for the liquid crystal director, the volume of the liquid crystal director is treated as an effective sphere with diameter $d$, and thereby the director will suffer a dissipative torque $\kappa=\pi\gamma d^{3}$ by Stokes relation.
An arranged Langevin equation corresponding to Equation (\ref{solvn}) is then obtained:
\begin{eqnarray}
\gamma\frac{\partial }{\partial t}n_{i}&=&-\left[ f_{g}\right]_{n_{i}}+\frac{\sigma\zeta}{\pi d^{3}} \nonumber \\
&\equiv &-\left[ f_{g}\right]_{n_{i}}+\xi, \label{lageq}
\end{eqnarray}
Details for this part will be discussed in
the next section. Moreover, the equation of motion for common use
\cite{Q1} can be estimated by considering the statisticly steady director
$\langle n_{i}\rangle$, i.e. ignore the (temperature-dependent) thermal fluctuations $\xi$
\begin{equation}
\gamma\frac{\partial }{\partial t}\langle n_{i}\rangle=-\left[
f_{g}\right] _{\langle n_{i}\rangle}, \label{lceq}
\end{equation}
with an effective order parameter $0<S<1$ in $\left[
f_{g}\right] _{\langle n_{i}\rangle}$.
Note that the function of the Lagrange multiplier $\lambda$ has been replaced here by
re-normalizing the director
$\left\vert \vec{n}\right\vert=1$ at each time step \cite{Q1}.
\section{Spatiotemporal Correlated Fluctuation Functions}
Based on the above-mentioned Langevin dynamics for liquid-crystal systems
 as the Equation (\ref{lageq}), we now detail
the spatiotemporal-correlated stochastic torque (fluctuation) $\xi$ as well as the corresponding dissipations
for the liquid-crystal directors. Mainly we consider
the Ornstein-Uhlenbeck process in 1+1 dimensions \cite{OU1} to generate the
spatiotemporal-correlated stochastic torque $\xi(z,t)$ with:
\begin{eqnarray}
\langle\xi(z,t)\rangle &=&0 \label{OUfluavg} \\
\langle\xi(z,t)\xi(z_0,t_0)\rangle &=&\frac{\gamma k_{B}T}{2\pi
d^{2}\lambda_{z}\tau}e^{-\frac{\left\vert t-t_{0}\right\vert
}{\tau}}e^{-\frac{\left\vert z-z_{0}\right\vert }{\lambda_{z}}} \label{OUflu}
\end{eqnarray}
Here, $\lambda_{z}$ and $\tau$ are the characteristic correlation
lengths in the spatial ($z$) and temporal ($t$) dimensions, respectively. $k_{B}$ is the
Boltzmann constant with a value of $1.38\times 10^{-23}$ $NmK^{-1}$,
$d$ is the effective size of the liquid crystal directors, and $T$
is the temperature of thermal environment. The Equation (\ref{OUflu}) can be further understood
by its power spectrum
$P\left(z_{0},\omega\right)=\frac{\gamma k_{B}T}{\pi
d^{2}\lambda_{z}}\frac{1}{1+\tau^{2}\omega ^{2}}$ which results from
the temporal Fourier transform of the correlation function
$\langle\xi(z_0,t)\xi(z_0,0)\rangle$. It is indicated that a longer correlation time
$\tau$ suppresses the higher-frequency (higher-energy) energy exchanges
and leads to a smaller fluctuation amplitude $[\frac{\gamma
k_{B}T}{2\pi d^{2}\lambda_{z}}\frac{1}{\tau}]^{\frac{1}{2}}$.
Likewise for the spatial Fourier transform of
$\langle\xi(z,t_0)\xi(0,t_0)\rangle$, i.e.
$P\left(k,t_0\right)=\frac{\gamma k_{B}T}{\pi
d^{2}\tau}\frac{1}{1+\lambda_{z}^{2}k^{2}}$, a longer correlation
length $\lambda_{z}$ implies the smaller-momentum (longer-wavelength)
energy exchanges, and causes a longer-ranged but smaller fluctuation
strength $[\frac{\gamma k_{B}T}{2\pi
d^{2}\tau}\frac{1}{\lambda_{z}}]^{\frac{1}{2}}$.

To generate this real-space fluctuation function $\xi(z,t)$ to
evaluate the motion of liquid crystals in Equation (\ref{lageq}), an exact
recursion relation with temporal spacing $\triangle_t$ and spatial
spacing $\triangle_z$ in the finite difference
method \cite{OU1} is applied:

\begin{eqnarray}
\xi \left( z,t\right)  &=&-\exp \left( -\frac{\triangle _{z}}{\lambda _{z}}-%
\frac{\triangle _{t}}{\tau }\right) \xi \left( z-\triangle
_{z},t-\triangle
_{t}\right)   \nonumber \\
&&+\exp \left( -\frac{\triangle _{z}}{\lambda _{z}}\right) \xi
\left(
z-\triangle _{z},t\right)   \nonumber \\
&&+\exp \left( -\frac{\triangle _{t}}{\tau }\right) \xi \left(
z,t-\triangle
_{t}\right)   \nonumber \\
&&+\left[ 1-\exp \left( -\frac{2\triangle _{z}}{\lambda _{z}}\right)
\right]
^{1/2}  \nonumber \\
&&\times \left[ 1-\exp \left( -\frac{2\triangle _{t}}{\tau }\right)
\right] ^{1/2}A\eta \left( z,t\right) \label{flufun}
\end{eqnarray}
Here, $\eta \left( z,t\right)$ is spatiotemporal Gaussian white
noise with $\langle
\eta(z,t)\eta(z',t')\rangle=\delta_{z,z'}\delta_{t,t'}$, and $A$ is
the fluctuation amplitude $[\frac{\gamma k_{B}T}{2\pi
d^{2}\tau\lambda_{z}}]^\frac{1}{2}$. Due to the exponentially
decaying correlation for spatiotemporal parameters and the characteristic
of white noise function $\langle\eta(z,t)\rangle=0$ in Equation (\ref{flufun}),
$\xi(z,t)$ is independent of its initial value and evidently meets
the condition in Equation (\ref{OUfluavg}). For the condition of fluctuation
correlations $\langle\xi(z,t)\xi(z_0,t_0)\rangle$ in Equation (\ref{OUflu}), numerical
statistics averaged over $10^3$ iterations
(each iteration includes the data number $N=100$ in spatial/temporal
space) are fulfilled to verify the validity of Equation (\ref{flufun}). The numerical results
by Equation (\ref{flufun}) are shown to compare well with the analytical results of Equation
(\ref{OUflu}) in Figures 1(b) and 2(b) for temporal and spatial arguments,
respectively. Figure 1(a) shows $\xi(z_0,t)$ of real temporal space
in one of the iterations and characterizes a $\tau$-period fluctuation
function. Figure 2(a) then presents $\xi(z,t_0)$ of real spatial
space with coherent length $\lambda_z$.

\begin{figure}[tbp]
\begin{center}
\includegraphics[scale=0.7]{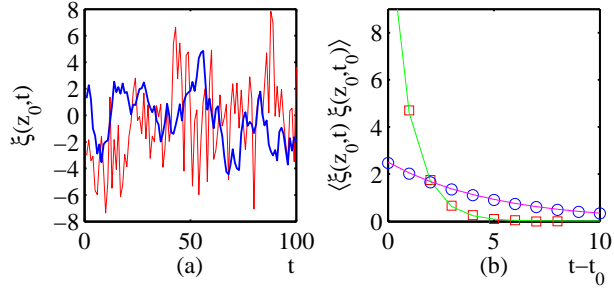}
\end{center}
\caption{Analyses of numerical fluctuation function $\xi$ for fixed
spatial state: $z=z_0$ and $\lambda_{z}=2\triangle_z$. (a)
Fluctuation function in real temporal space for $\tau=\triangle_t$
(thin red line) and $\tau=5\triangle_t$ (bold blue line). (b)
Temporal correlation function for $\tau=\triangle_t$ (symbol $\Box$)
and $\tau=5\triangle_t$ (symbol $\bigcirc$). Solid lines indicate
the corresponding analytical results.} \label{fix_x}
\end{figure}

\begin{figure}[tbp]
\begin{center}
\includegraphics[scale=0.7]{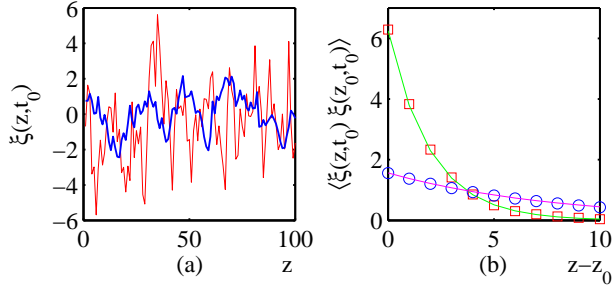}
\end{center}
\caption{Analyses of numerical fluctuation function $\xi$ for fixed
temporal state: $t=t_0$ and $\tau=2\triangle_t$. (a) Fluctuation
function in real spatial space for $\lambda_z=2\triangle_z$ (thin red
line) and $\lambda_z=8\triangle_z$ (bold blue line). (b) Spatial
correlation function for $\lambda_z=2\triangle_z$ (symbol $\Box$)
and $\lambda_z=8\triangle_z$ (symbol $\bigcirc$). Solid lines
indicate the corresponding analytical results.} \label{fix_t}
\end{figure}

Implemented with Equation (\ref{flufun}) for the real-space fluctuation
function $\xi(z,t)$, an updated formula by discretizing Equation
(\ref{lageq}) can be obtained:
\begin{eqnarray}
n_{i}(z,t+\Delta t)-n_{i}(z,t)&=&\frac{\Delta t}{\gamma}(-\left[
f_{g}\right] _{n_{i}}+\xi(z,t)), \text{ \ \ } i\in\{x,y,z\} \label{eq_fd}
\end{eqnarray}
for the equation of directors in 1+1 dimension systems. This
algorithm includes the influences of thermal baths and supplies a
theoretical foundation for non-equilibrium liquid crystal dynamics.

\section{Numerical Analyses of Non-equilibrium Dynamics}
\begin{figure}[tbp]
\begin{center}
\includegraphics[scale=0.6]{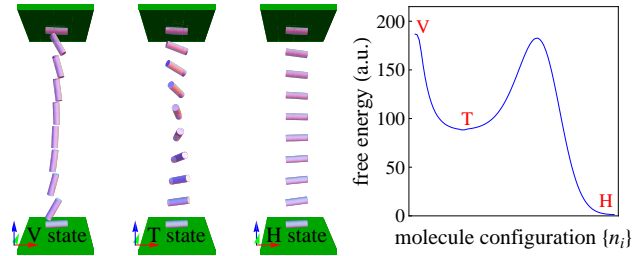}
\end{center}
\caption{Liquid crystal configurations and the corresponding Gibbs
free energy in $\pi$ configuration: bend ($V$), twist ($T$), and homogenous
($H$) equilibrium states.} \label{OCB}
\end{figure}
To concretely understand the influence of the thermal fluctuations on the
liquid-crystal metastable systems, we now consider an actual case with $\pi$ configuration \cite{OCB1}.
 As shown in
Figure 3, with the Frank-Oseen elastic theory, bend ($V$), twist ($T$), and homogenous ($H$)
states form a metastable status. A transition barrier exists
between the $T$ and $H$ states, leading to a local minimum
localization on the $T$ state for the time evolution from the $V$ state. (Here the red,
green, and blue arrows indicate the spatial $\hat{x}$, $\hat{y}$,
and $\hat{z}$ directions, respectively.)
In this work, we mainly explain how the initial $V$ state can transit through the
locally-stable $T$ state, and finally into the globally-stable $H$ in the long run
as experimental observations.
To statistically evaluate the influences of the fluctuation and
dissipation, we consider an ensemble ($10^3$ iterations) of
liquid crystal cells for each set-up conditions of parameters.
The associated parameters are defined as: $\gamma=50 mPa\cdot
\sec $, $K_{11}=13.6 pN$, $K_{22}=10.0 pN$, $K_{33}=14.7 pN$,
$\varepsilon _{\parallel }=10$, $\varepsilon _{\perp }=5$, and $T=300K$.
The thickness of the cell is set as $4.5$ $\mu m$. No
chiral agent is added. The pre-tilt angle between the liquid crystal
directors and the alignment (green) layers is set to $5^{\circ }$.
For the finite difference algorithm, the spatial grid is
$\triangle_{z}=0.15 \mu m$ and the temporal grid is
$\triangle_{t}=0.01 ms$. Besides, the effective director size $d$ is treated as about
$10^{1}-10^{2}nm$ for this study, which is $1-2$ order larger than the liquid crystal molecules.
For each iteration of the ensemble, the $V$ state is initialized
by applying a bias $15V$, which lasts for a long enough period
($200ms$ in our cases), and the voltage is turned off at $t\geq 0$
in order to observe the evolution of the liquid crystals.
Figure 4(a) illustrates the transition rate
$\Gamma_{H}=1/\langle t_{V\rightarrow H}\rangle$ versus fluctuation
amplitude $A$, where $\langle t_{V\rightarrow H}\rangle$ represents
the average time of $V\rightarrow H$ transitions of
the ensemble. In the Figure 4(a),
in the absence of the thermal bath (at $A=0Pa$),
the process of minimizing-energy as the Equation (\ref{lceq}) drives the dynamics of
the liquid crystals, and consequently the initial $V$ states in all of the 1000 iterations are found to
follow the free-energy-decreasing trace into the locally-stable $T$ state as the blue solid-line segment
form $V$ to $T$ in the Figure 4(b). No iteration of transition from $V$ to $H$ is found in our simulation in this case.
This results in the infinite transition time $\langle t_{V\rightarrow H}\rangle=\infty$ for $V$ to $H$ state,
or the zero transition rate
$\Gamma_{H}=1/\langle t_{V\rightarrow H}\rangle=0$ as in Figure 4(a).
(Note that Figure 4(a) shows the log-scale plot for clear appearance, and the infinite time is realized by
a cur-off time $10^5ms$ in real numerical calculations.)
As the fluctuation amplitude $A$ increases ($A\neq 0Pa$), the liquid crystals contain not only the
elastic free energy but also the finite fluctuation energy; Or alternatively, the liquid crystals
suffer an extra stochastic torque $\xi$ in addition to the fundamental elastic torques. This allows
the initial $V$ state in some/few iterations of the 1000 iterations have the possibility to deviate from the free-energy-decreasing trace, and are found to proceed another non-equilibrium path (as the green dashed curve in Figure 4(b)) to reach the final $H$ states,
although the paths of the most iterations still remain on the
free-energy-decreasing trace (red dashed-dotted curve in Figure 4(b)) to be localized at $T$ states in our calculations.
This causes the transition time $\langle t_{V\rightarrow H}\rangle$ of some of the 1000 iterations to be finite and thereby the finite statistic
transition rate $\Gamma_{H}=1/\langle t_{V\rightarrow H}\rangle$ as in Figure 4(a).
Consequently, it is found that larger
fluctuation amplitude $A$ can cause more iterations of the ensemble to proceed the non-equilibrium path
($V\rightarrow H$; green dashed curve in Figure 4(b)) and raise the transition rate
$\Gamma_{H}=1/\langle t_{V\rightarrow H}\rangle$ as in Figure 4(a).
Here, Figure 4(b) illustrates quantitatively the energy evolution traces of the
transitions $V\rightarrow T$ and $V\rightarrow H$ for $A=100$ $Pa$, $\tau=\triangle_t$, and
$\lambda_z=\triangle_z$. For the large fluctuation amplitude ($A\rightarrow\infty$ $Pa$),
however, the overwhelming fluctuation torque leads
to strong-fluctuated liquid crystal orientations, and thereby the
liquid crystal configurations ($V$, $T$, and $H$) cannot be well
identified. This is far away from phenomenological
observations and is beyond the scope of this work.
In addition to these analyses of the transition dynamics on the aspect of time and energy as in Figures 4(a) and 4(b),
respectively, We further analyze the corresponding time evolutions of the directors
to help to understand the transition dynamics.

Figure 5 purposes to briefly characterize the liquid-crystal profiles of the above-mentioned transitions first,
and the detailed time-evolution traces of the corresponding liquid-crystal directors are then depicted in Figure 6.
The characteristic director profiles of the transition $V\rightarrow T$, which corresponds to the red dashed-dotted curve on Figure 4(b), is characterized on Figure 5(a). The characteristic director profiles of the transition $V\rightarrow H$, which corresponds to the green dashed curve on Figure 4(b), are found to
have two possible time-evolution traces in our simulations and
are characterized by $S_{0}$ and $S_{1}$ in Figures 5(b) and 5(c), respectively.
In Figure 6(a) (corresponding to Figure 5(a)) the orientations of the individual liquid-crystal directors
draws the bend ($V$) state at $t=0$ and denoted by ($n_x,n_y$)$\rightarrow$($0,0$),
i.e. $|n_z|\rightarrow\sqrt{1-n_x^2-n_y^2}=1$. By the time going, these directors
then proceed (along the red dashed-dotted energy trace in Figure 4(b)) to the locally-stable $T$ state, and consequently
describe the transition $V\rightarrow T$ over the long run. Here, an alternative
anti-clockwise-twist profiles is analogous to the presented clockwise-twist
one, and is ignored in the context. Similar anti-clockwise/clockwise configurations occurred on the transition $V\rightarrow H$ is ignored below as well.
Figures 6(b) and 6(c) numerically show two possible time-evolution traces of directors
($n_x,n_y$) corresponding to the transitions $V \rightarrow S_{0} \rightarrow H$
and $V \rightarrow S_{1} \rightarrow H$, respectively. Note these two transitions ($V \rightarrow S_{0} \rightarrow H$
and $V \rightarrow S_{1} \rightarrow H$) show an indistinguishable energy trace as the green dashed curve in Figure 4(b) in our simulations. All these construct the fundamental understandings to help to
study the complete $V \rightarrow H$ transition in real three-dimensional liquid-crystal $\pi$ systems
as below.

\begin{figure}[tbp]
\begin{center}
\includegraphics[scale=0.71]{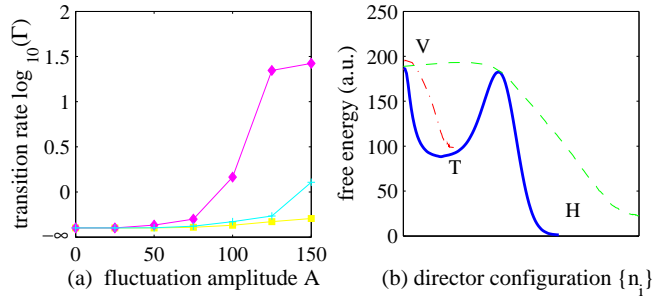}
\end{center}
\caption{(a) Transition rate of $V\rightarrow H$ as a function of fluctuation amplitude $A$
with characteristic correlation scale $\tau=\triangle_t$, $\lambda_z=\triangle_z$ (mark $\Box$),
$\tau=\triangle_t$, $\lambda_z=5\triangle_z$ (mark $+$) , and
$\tau=5\triangle_t$, $\lambda_z=\triangle_z$ (mark $\lozenge$).
In which the horizontal axis corresponds to the $d=93nm$, $47nm$, and $31nm$
for $A=50$, $100$, and $150Pa$, respectively for $\tau =\Delta t$, $\lambda
_{z}=\Delta z$, and $T=300K$. And for the case of $\tau =\Delta t$, $\lambda
_{z}=5\Delta z$ and $\tau =5\Delta t$, $\lambda _{z}=\Delta z$, $A=50$, $100$%
, and $150Pa$ corresponds to $d=41nm$, $20nm$, and $14nm$, respectively.
(b) Energy evolution traces of alternative non-equilibrium transitions for $V\rightarrow T$ (dashed-dotted curve) and
$V\rightarrow H$ (dashed curve) for $A=100$ $Pa$ ($T=300$ $K$ with $d=50$ $nm$), $\tau=\triangle_t$, and $\lambda_z=\triangle_z$. } \label{rate}
\end{figure}

\begin{figure}[tbp]
\begin{center}
\includegraphics[scale=0.6]{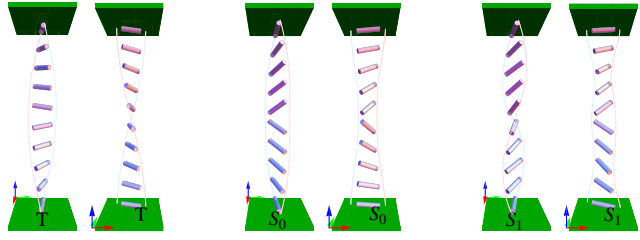}
\end{center}
\caption{Characteristics of liquid crystal configurations of alternative non-equilibrium
transitions for (a) $V\rightarrow T$ process, (b) $V \rightarrow S_{0} \rightarrow H$ process, (c)
$V \rightarrow S_{1} \rightarrow H$ process viewed from $+x$ (left) and $-y$ (right).
 } \label{OCBms}
\end{figure}

\begin{figure}[tbp]
\begin{center}
\includegraphics[scale=0.8]{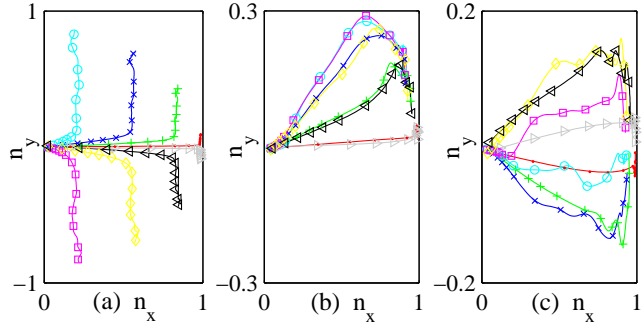}
\end{center}
\caption{Evolution traces of liquid crystal director ($n_x,n_y$)
corresponding to (a) $V\rightarrow T$ process, (b) $V \rightarrow S_{0} \rightarrow H$ process, (c)
$V \rightarrow S_{1} \rightarrow H$ process. In which the directors at $z=d/9$ (mark $\bullet$),
$z=2d/9$ (mark $+$), $z=3d/9$ (mark $\times$), $z=4d/9$ (mark
$\circ$), $z=5d/9$ (mark $\square$), $z=6d/9$ (mark $\lozenge$),
$z=7d/9$ (mark $\vartriangleleft$), $z=8d/9$ (mark
$\vartriangleright$) are shown. } \label{trace_all}
\end{figure}
\begin{figure}[tbp]
\begin{center}
\includegraphics[scale=0.45]{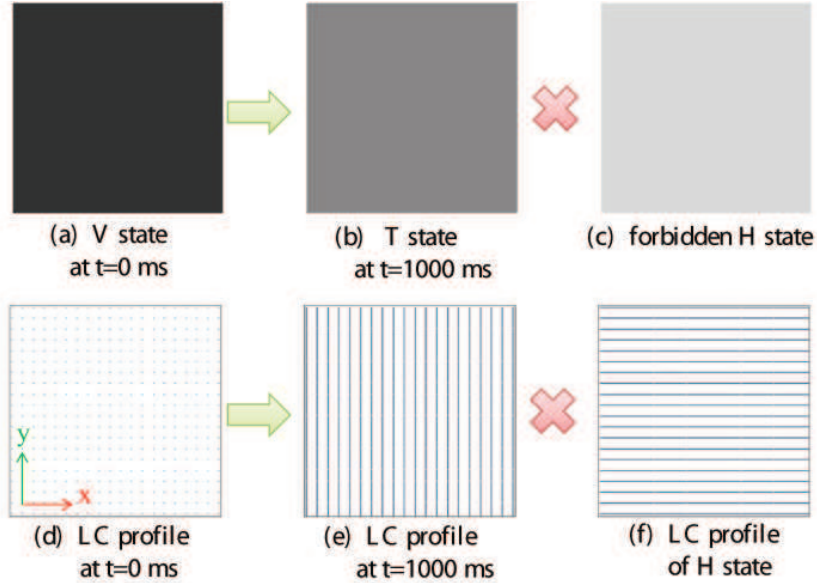}
\end{center}
\caption{ Optical transmittances of (a) the initial $V$ state at
$t=0$ $ms$, (b) the stable $T$ state at $t=1000$ $ms$, and (c) the
forbidden $H$ state for the $V\rightarrow T,H$ transition without
thermal coupling $A'=0$. The corresponding liquid crystal directors
(LC profiles) at $z=d/2$ are showed as below plots (d), (e), and
(f), respectively. } \label{noT}
\end{figure}
\begin{figure}[tbp]
\begin{center}
\includegraphics[scale=0.45]{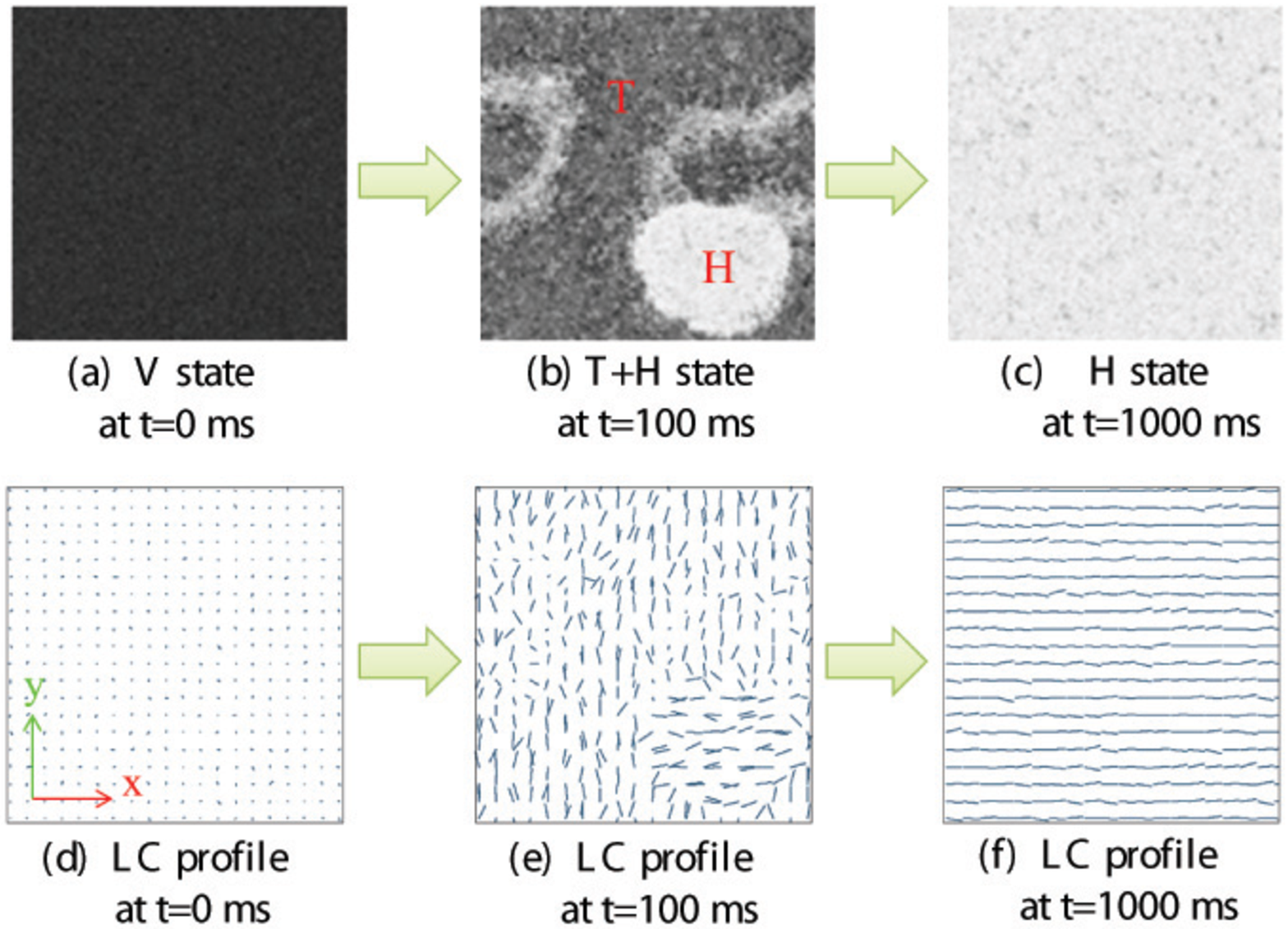}
\end{center}
\caption{ Optical transmittances of (a) the initial $V$ state at
$t=0$ $ms$, (b) the transient $T+H$ state at $t=100$ $ms$, and (c)
the stable $H$ state at $t=1000$ $ms$ for the $V\rightarrow T,H$
transition with thermal coupling $A'=100$. The corresponding liquid
crystal directors (LC profiles) at $z=d/2$ are showed as below plots
(d), (e), and (f), respectively.  } \label{T}
\end{figure}
\begin{figure}[tbp]
\begin{center}
\includegraphics[scale=0.325]{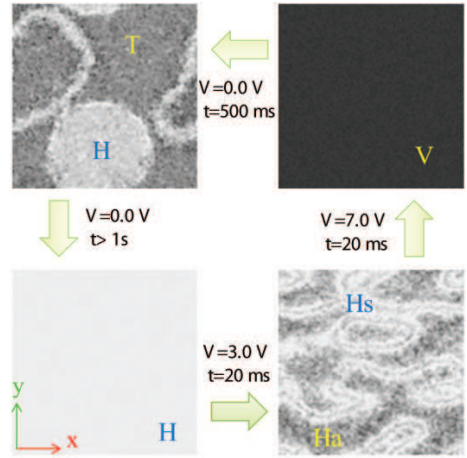}
\end{center}
\caption{Transition processes of $\pi$-configuration under the bias sequence.
In the field-on period, the homogenous (H) state transforms to the
stochastic symmetric (Hs) and asymmetric (Ha) states, and further proceed to
the bend (V) state under higher biases. In the field-off period, the bend (V) state
first relaxes to the twist (T) and homogenous (H) states as described in the context, and recovers to the overall homogenous (H) configuration in the long run.} \label{OCBphase}
\end{figure}

In the following section, we numerically investigate an extension study of the three-dimensional
liquid-crystal metastable system to give a
result corresponding to the actual experimental observations \cite{OCB1}.
It is emphasized that the spatial/temporal grid size ($\Delta_{z}$
and $\Delta_{t}$) is herein assumed to be much larger than the
coherent length/time ($\lambda_{z}$ and $\tau$), such that the
fluctuations for each time step at the individual site are almost
independent, i.e. $\xi \left(z,t\right)=A\eta \left( z,t\right)$ in
Equation (\ref{flufun}). This thereby allows a straightforward extension to the three-dimensional
condition $\xi \left(x,y,z,t\right)=A'\eta \left( x,y,z,t\right)$
for the studied system. The relevant parameters of the $\pi$
configuration are similar to the above-mentioned, except for
$\Delta_{z}=0.45$ $\mu m$ as well as the considerations of optical
transmittances \cite{aOptic} related to practical observations. The associated optical parameters
are set as: the incident wavelength is $0.55$ $\mu m$,
ordinary refractive index of liquid crystals is $n_{o}=1.5$,
extraordinary refractive index of liquid crystals is $n_{e}=1.58 $,
and the absorption axes of the upper
and lower polarizers are located at $\phi =45^{\circ }$, and
$135^{\circ }$ in the $x-y$ plane, respectively. The
numerical results for the metastable transition $V\rightarrow T,H$
without ($A'=0$) and with ($A'=100$) thermal couplings are shown in
Figures \ref{noT} and \ref{T}, respectively. In the case of $A'=0$
in Figure \ref{noT}, the initial $V$ state follows the
energy-minimizing process in the elastic continuum theory and leads
to the localization at the $T$ state as described above.
However, the observable $H$ state exhibits significant differences
of liquid crystal configurations to the $T$ state in Figures
\ref{noT}(e) and \ref{noT}(f) and is forbidden by the energy barrier
(Figure \ref{rate}(b)). In the case of $A'=100$ in Figure \ref{T},
thermal fluctuations as well as the associated correlations offer
the possibility of the non-equilibrium $V\rightarrow H$ transitions
and cause the $T+H$ transient states in $100$ $ms$ after the bias is
dropped to zero (Figure \ref{T}(b)). Thereafter, the elastic dynamics describing the
disclination motions \cite{OCB1} drives a final $H$ stable state as in Figure
\ref{T}(c). This gives an explanation for the overall $V\rightarrow H$
process. Overall, with thermal fluctuations, a straightforward analysis
corresponding to the observed state transitions of $\pi$-configuration
in Reference \cite{OCB1} can be
realized as in Figure (\ref{OCBphase}).

\section{Conclusions}

In this work, based on the elastic continuum theory, we have
introduced the Langevin algorithm with the Ornstein-Uhlenbeck
process to realize the non-equilibrium dynamics of liquid crystal
systems coupled weakly to the thermal environment. In which the
Ornstein-Uhlenbeck process supplies a connection to the natural
spatial and temporal correlation scales of thermal fluctuations.
All these allow one to simulate a non-equilibrium time evolution and
give physical insights into the processes taking place in the
studied systems. It is noted that another thermal hopping (
hop over the potential barrier, i.e. a direct transition $T\rightarrow H$ in
Figure \ref{rate}(b)) dynamics can be set up into this system through
strong coupling cases, but go beyond this study's main goal.
Definitely, this work provides one possible explanation for the complete transition
$V\rightarrow H$ in the $\pi$
configurations and supplies the corresponding real-time simulation results
for the experimental observations \cite{OCB1}.

\appendix
\section{Fluctuation-induced Vortex/anitvortex in VA cells}
\begin{figure}[tbp]
\begin{center}
\includegraphics[scale=0.425]{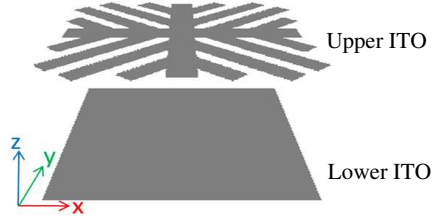}
\end{center}
\caption{Structures of the VA liquid crystal cells with patterned ITO (indium tin oxide) electrodes. The cell volume is set to be $50um\times50um\times4um$ in width$\times$length$\times$height ($x\times y\times z$) dimensions.  } \label{VAstr1}
\end{figure}

\begin{figure}[tbp]
\begin{center}
\includegraphics[scale=0.475]{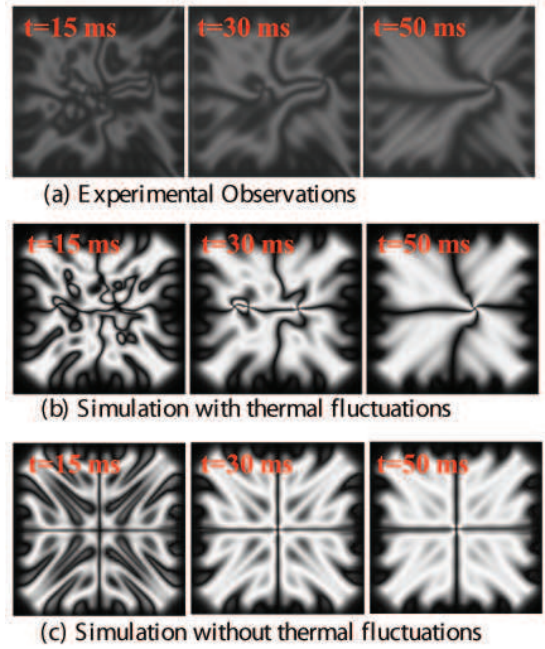}
\end{center}
\caption{Optical transmittances of the VA cell under bias 5V in (a) experimental observations,
(b) numerical analyses with fluctuations, and (c) numerical analyses without fluctuations. In which
(a) and (b) show the comparable
emergence and annihilation behaviors of vortex/antivortex by time. (c) shows the regular time evolutions
of LC profiles on transmittances.} \label{vortex1}
\end{figure}

In this appendix, we briefly show the fluctuation-induced (stochastic) vortex/antivortex through experimental and theoretical works, and the associated fluctuation amplitude $A'$ can be estimated as well. The used structure of the VA (vertical alignment) liquid crystal cells is depicted in Figure (\ref{VAstr1}). The associated LC parameters are defined as: $\gamma=135 mPa\cdot
\sec $, $K_{11}=13.6 pN$, $K_{22}=10.0 pN$, $K_{33}=14.7 pN$,
$\varepsilon _{\parallel }=3.6$, and $\varepsilon _{\perp }=7.3$. Here, the fluctuation amplitude $A'$ are decided by fitting the density and period of vortexes/antivortexes with experimental observations, and obtained as $A'=80-125 Pa$. These amplitude value correspond to the fluctuated angle of the director around $5^{\circ }$ for studied cases in our simulations.
The pre-tilt angle between the liquid crystal
directors and the alignment (xy) plane is set to $90^{\circ }$ as the experimental setup. For optical analyses, the
incident light wavelength $\lambda$ is set as $0.55$ $\mu m$,
$n_{o}=1.5$, $ n_{e}=1.6 $, and the absorption axes of the upper
and lower polarizers are located at $\phi =0^{\circ }$, and
$90^{\circ }$ in the $x-y$ plane, respectively. Besides, a long enough pre-simulation (lasting 100 ms) is fulfilled to simulate the natural/initial LC status for the sequential numerical simulation. Numerical results as well as experimental observations for optical transmittances are illustrated in Figure \ref{vortex1} by time. It indicated that the thermal fluctuations can
introduce stochastic vortex/antivortex as observed in experiments \ref{vortex1}(a), while without fluctuations a regular time evolution of LC profiles is predicted as in Figure \ref{vortex1}(c). Definitely, these allow a further analysis on the emergence and annihilation behaviors of vortex/antivortex associated the spatial and temporal correlations of fluctuations \cite{fl1}, while beyond the scope of this work.

\section{Effects of Fluctuations on LC response time}
\begin{figure}[tbp]
\begin{center}
\includegraphics[scale=0.425]{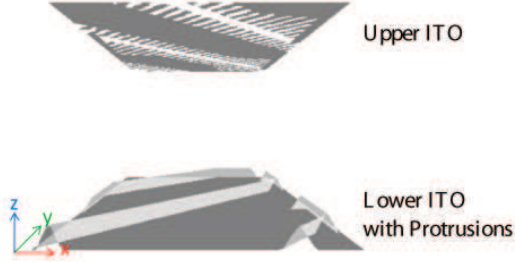}
\end{center}
\caption{Structures of the MVA liquid crystal cells, in which the upper ITO has fine-slit patterns and the lower ITO stacks with protrusion layers. The cell volume is set to be $160um\times200um\times4um$ in width$\times$length$\times$height ($x\times y\times z$) dimensions.} \label{VAstr2}
\end{figure}

\begin{figure}[tbp]
\begin{center}
\includegraphics[scale=0.375]{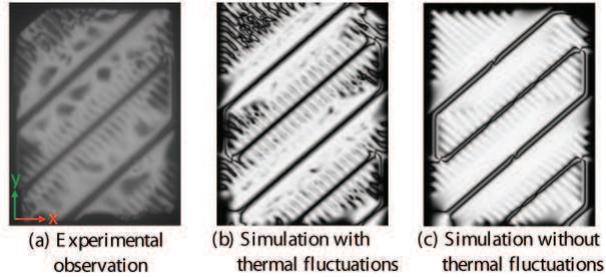}
\end{center}
\caption{Optical transmittances of the MVA cell with bias 5 V at t=16 ms by (a) experimental observations,
(b) numerical simulations with fluctuations, and (c) numerical simulations without fluctuations. In which
(a) and (b) show the stochastic vortex/antivortex spots. (c) shows the regular transmittance pattern.} \label{vortex2}
\end{figure}

\begin{figure}[tbp]
\begin{center}
\includegraphics[scale=0.55]{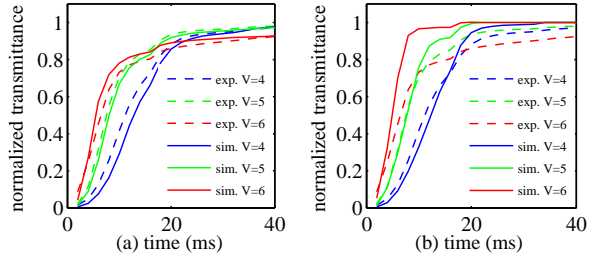}
\end{center}
\caption{Normalized transmittances as a function of time at biases V=4,5,6 V for (a) numerical simulations with fluctuations, and (b) numerical simulations without fluctuations. The experimental results are added to both analyses for comparison. } \label{RTime}
\end{figure}

In this appendix, not only the fluctuation-induced vortex/antivortex as described at Appendix A, the thermal fluctuations are shown to have appreciable influences on the LC response behaviors (against biases). In the meantime, this influence supplies an alternative method to quantitatively evaluate the fluctuation amplitude $A'$. The used structure of the MVA (multi-domain vertical alignment) liquid crystal cells is depicted in Figure (\ref{VAstr2}), in which the upper ITO (indium tin oxide) has fine-slit patterns and the lower ITO stacks with protrusion layers for the tilted-alignment purposes. The associated parameters are referred to these at Appendix A; while the fluctuation amplitude $A'$ are decided to be 115 $Pa$ by fitting the response time as below. Numerical results as well as the experimental observations for the optical transmittance at t=16 ms and V=5 V are shown in Figure \ref{vortex2}.
Normalized transmittances as a function of time at biases V=4, 5, and 6 V are then depicted by numerical simulations with fluctuations in Figure \ref{RTime}(a), numerical simulations without fluctuations Figure \ref{RTime}(b), and experimental observations in Figures \ref{RTime}(a) and (b) for comparison. Compared to the common understandings: larger bias/field causes faster LC response in field-on process as shown in Figure \ref{RTime}(b) (solid curves; without fluctuations), thermal fluctuations as well as experimental observations give another story herein. Besides fast LC response in field-on process, large bias/field cause simultaneously dense vortex spots as shown in Figure \ref{vortex2}(a). These observations agree with the numerical results if the fluctuations are included as Figure \ref{vortex2}(b). Definitely, the sequential reliefs of the entangled vortexes could cost more time to approach the final stable status, and thereby could reduce the liquid crystal response time. This can be observed in Figure \ref{RTime}(a): compared between these curves with V=4 and 5 V, the response time with larger bias (5V) is indeed faster than that with lower bias (4V); however, compared between these curves with V=5 and 6V, the response time with larger bias (6V) is slower than that with lower bias (5V). In Figure \ref{RTime}(b), however, the numerical simulations without fluctuations could not help to understand these experimental observations. Substantially, this gives an example to indicate the significance of fluctuations in some real liquid crystal systems.

\end{document}